\documentclass[11pt]{article}
\usepackage[dvips]{graphicx}
\usepackage{graphics}
\usepackage{graphicx}
\begin{document}
\begin{titlepage}
\title{\bf Quantum Protectorate  and Microscopic Models of
Magnetism\thanks{International Journal of Modern Physics B16, N 5,
(2002)pp.803-823} }
\author{
A.L.Kuzemsky \thanks {E-mail:kuzemsky@thsun1.jinr.ru;
http://thsun1.jinr.ru/~kuzemsky}
\\
{\it Bogoliubov Laboratory of Theoretical Physics,} \\
{\it  Joint Institute for Nuclear Research,}\\
{\it 141980 Dubna, Moscow Region, Russia.}}
\date{}
\maketitle
\begin{abstract}
Some   physical implications involved in a new concept, termed the
"quantum protectorate" (QP), are developed and discussed. This is
done by considering the idea of quantum protectorate  in the
context of quantum theory of magnetism. It is suggested that the
difficulties in the formulation of   quantum theory of magnetism
at the  microscopic level,  that are related to the choice of
relevant models,  can be understood better in the light of the QP
concept . We argue that the difficulties in the formulation of
adequate microscopic models of electron and magnetic properties of
materials are intimately related to dual, itinerant and localized
behaviour of electrons. We formulate a criterion of what basic
picture describes best  this dual behaviour. The main suggestion
is that quasi-particle excitation spectra might provide
distinctive signatures and good criteria for the appropriate
choice of the relevant model.
\end{abstract}
\end{titlepage}
\section{Introduction}
It is well known that there are many branches of physics and
chemistry where phenomena occur which cannot be described in the
framework of   interactions amongst a few particles\cite{kohn}.
As a rule, these phenomena arise essentially from the cooperative
behaviour of a large number of   particles. Such many-body
problems are of   great interest not only because of the nature
of phenomena themselves, but also because of the intrinsic
difficulty in solving problems which involve   interactions of
many particles  in terms of   known Anderson  statement that "more
is different"~\cite{and1}. It is often difficult to formulate a
fully consistent and adequate microscopic theory of complex
cooperative phenomena. In ref.\cite{pin1},  the  authors invented
an idea of a quantum protectorate, "a stable state of matter,
whose generic low-energy properties are determined by a
higher-organizing principle and nothing else"\cite{pin1}. This
idea brings into physics the concept that reminds   the
uncertainty relations of quantum mechanics . The notion of QP was
introduced to unify some generic features of complex physical
systems on different energy scales,  and is a certain
reformulation of the conservation laws and symmetry breaking
concepts\cite{ander}. As typical examples of QP, the crystalline
state, the Landau fermi liquid, the state of matter represented
by conventional metals and normal $ ^3He $ ({\it
cf.}\cite{volov1},\cite{volov2}) , and the quantum Hall effect
were considered. The sources of quantum protection in high-$T_c$
superconductivity and low-dimensional systems were discussed in
refs.\cite{and2}-\cite{pin2}. According to Anderson\cite{and2},
"the source of quantum protection is likely to be a collective
state of the quantum field, in which the individual particles are
sufficiently tightly coupled that elementary excitations no
longer involve just a few particles, but are collective
excitations of the whole system. As a result, macroscopic
behaviour is mostly determined by overall conservation laws". In
the same manner the concept of a spontaneous breakdown of
symmetry enters through the observation that the symmetry of a
physical system could be lower than the symmetry of  the basic
equations describing the system\cite{ander},\cite{call}. This
situation is encountered in non-relativistic statistical
mechanics. A typical example is provided by the formation of a
crystal which is not invariant under all space translations,
although the basic equations of equilibrium mechanics are. In
this article, I will attempt to relate the term of a  quantum
protectorate  and the foundations of   quantum theory of
magnetism. I will not touch the low-dimensional systems that were
discussed already comprehensively in
refs.\cite{and2}-\cite{pin2}. I concentrate on the problem of
choosing the most adequate microscopic model of magnetism of
materials and, in particular, related to the duality of localized
and itinerant behaviour of electrons where the microscopic theory
meets the most serious difficulties. To justify this statement
and to introduce all necessary notions that are relevant for the
present discussion, we very briefly recall the basic facts of the
microscopic approach to magnetism.
\section{ Magnetic Degrees of Freedom }
The discussion in this paper is concentrated on the right
definition of the fundamental "magnetic" degrees of freedom and
their correct model description for complex magnetic systems. We
shall first describe   the phenomenology of the magnetic
materials to look at the physics involved. The problem of
identification of the fundamental "magnetic" degrees of freedom
in complex materials is rather nontrivial. Let us discuss
briefly, to give a flavor only, the very intriguing problem of
the electron dual behaviour. The existence and properties of
localized and itinerant magnetism in insulators, metals, oxides
and alloys and their interplay in complex materials is an
interesting and not yet fully understood problem of quantum
theory of magnetism\cite{kuz3},\cite{kuz7}. The central problem of
recent efforts is to investigate the interplay and competition of
the insulating, metallic, superconducting, and heavy fermion
behaviour versus the magnetic behaviour, especially in the
vicinity of a transition to a magnetically ordered state. The
behaviour and the true nature of the electronic and spin states
and their quasi-particle dynamics are of central importance  to
the understanding of the physics of strongly correlated systems
such as magnetism and metal-insulator transition  in metals and
oxides, heavy fermion states , superconductivity and their
competition with magnetism\cite{pin3}. The strongly correlated
electron systems are systems in which electron correlations
dominate. An important problem in understanding the physical
behaviour of these systems was the connection between relevant
underlying chemical, crystal and electronic structure, and the
magnetic and transport properties which continue to be the
subject of intensive debates\cite{mathur}. Strongly correlated
$d$ and $f$ electron systems are   of special interest\cite{kuz1}
. In these materials electron  correlation effects are essential
and, moreover, their spectra are complex, {\it i.e.}, have many
branches.   Importance of the studies on strongly correlated
electron systems are concerned with a fundamental problem of
electronic solid state theory, namely, with a tendency of $3(4)
d$ electrons in transition metals and compounds and  $4(5) f$
electrons in rare-earth metals and compounds and alloys to
exhibit both localized and delocalized behaviour
\cite{kuz3},\cite{kuz2}. Many electronic and magnetic features of
these substances relate intimately to this dual behaviour of the
relevant electronic states. For example, there are some alloy
systems in which radical changes in physical properties occur
with relatively modest changes in chemical composition or
structural perfection of the crystal lattice\cite{kuz1}.  Due to
competing interactions of comparable strength, more complex
ground states than usually supposed may be realized. The strong
correlation effects among electrons, which lead to the formation
of the heavy fermion state take part to some extent in formation
of a magnetically ordered phase, and thus imply that the very
delicate competition and interplay of interactions exist in these
substances\cite{mathu}. For most of the heavy fermion
superconductors, cooperative magnetism, usually some kind of
antiferromagnetic ordering was observed in the "vicinity" of
superconductivity. In the case of U-based compounds, the two
phenomena, antiferromagnetism and superconductivity coexist on a
microscopic scale, while they seem to compete with each other in
the Ce-based systems\cite{lonz1}. For a Kondo lattice system, the
formation of a Neel state via the RKKY intersite interaction
compete with the formation of a local Kondo singlet . Recent data
for many heavy fermion Ce- or U-based compounds and alloys
display a pronounced non-Fermi-liquid behaviour. A number of
theoretical scenarios have been proposed and they can be broadly
classified into  two categories which deal  with the localized
and extended states of $f$-electrons. Of special interest is the
unsolved controversial problem of the reduced magnetic moment in
Ce- and U-based alloys and the description of the heavy fermion
state in the presence of the coexisting magnetic state. In other
words, the main interest is in  the understanding of the
competition of intra-site (Kondo screening) and inter-site (RKKY
exchange) interactions. Depending on the relative magnitudes of
the Kondo and  RKKY scales, materials with different
characteristics are found which are classified as non-magnetic
and magnetic concentrated Kondo systems. The latter, "Kondo
magnets", are  of main interest\cite{kuz1}. Furthermore, there
are effects which have a very complicated and controversial
origin. There are some experimental evidences that peculiar
magnetism of some quasi-ternary heavy fermion alloys is not that
of  localized systems, but have some features of band magnetism.
Thus, in addition to the pronounced non-Fermi-liquid effects  in
thermodynamic and transport properties, the outstanding problems
include small magnetic moments and possible transitions from a
localized moment ordered phase to a kind of "heavy fermion band
magnet"\cite{allen} - \cite{steg2}. These features reflect the
very delicate interplay and competition of interactions  and
changes in a chemical composition. As a rule, very little
intuitive insight could be gained from this very complicated
behaviour. The QP is an umbrella term for a theoretical approach
which seems designed specifically to analyze such problems.
\section{ Microscopic Picture of Magnetism in Materials.}
In this Section we recall the foundations of the quantum theory of
magnetism in a sketchy form. Magnetism in materials such as iron
and nickel results from the cooperative alignment of the
microscopic magnetic moments of electrons in the material. The
interactions between the microscopic magnets are described
mathematically by the form of the Hamiltonian of the system. The
Hamiltonian depends on some parameters, or coupling constants,
which measure the strength of different kinds of interactions. The
magnetization, which is measured experimentally, is related to
the average or mean alignment of the microscopic magnets. It is
clear that some of the parameters describing the transition to
the magnetically ordered state do depend on the detailed nature
of the forces between the microscopic magnetic moments. The
strength of the interaction will be reflected in the critical
temperature which is high if the aligning forces are strong and
low if   they are weak. In quantum theory of magnetism, the
method of model Hamiltonians has proved to be very effective.
Without exaggeration, one can say that the great advances in the
physics of magnetic phenomena are to a considerable extent due to
the use of   very simplified and schematic model representations
for the theoretical
interpretation.\\
\subsection{ Heisenberg Model }
 The Heisenberg model is based on
the assumption that the wave functions of magnetically active
electrons in crystals differ little from the atomic orbitals. The
physical picture can be represented by a model in which the
localized magnetic moments originating from ions with incomplete
shells interact through a short-range interaction. Individual
spin moments form a regular lattice. The  model of a system of
spins on  a lattice is termed the Heisenberg
ferromagnet\cite{tyab} and establishes the origin of the coupling
constant as the exchange energy. The Heisenberg ferromagnet in  a
magnetic field $H$ is described by the Hamiltonian
\begin{equation}
\label{eq.1} H = -  \sum_{ij}  J(i-j) \vec S_{i} \vec S_{j}
-g\mu_{B}H\sum_{i}S_{i}^{z}
\end{equation}
The coupling coefficient $J(i-j)$ is the measure of the  exchange
interaction between  spins at the lattice sites $i$ and $j$ and
is defined usually to have the property J(i - j = 0) = 0. This
constraint means that only the inter-exchange interactions are
taken into account. The coupling, in principle, can be of a more
general type (non-Heisenberg terms). For crystal lattices in
which every ion is at the centre of symmetry, the exchange
parameter has the property $$ J(i-j) = J(j-i)$$  We can rewrite
then the Hamiltonian (\ref{eq.1}) as
\begin{equation}
\label{eq.2} H = -  \sum_{ij} J(i-j) ( S^z_{i}S^z_{j} +
 S^+_{i}S^-_{j})
\end{equation}
Here $S^{\pm} = S^x \pm iS^y$ are the raising and lowering spin
angular momentum operators. The complete set of spin commutation
relations is
\begin{eqnarray} \nonumber
[S^{+}_{i},S^{-}_{j}]_{-} = 2S^{z}_{i} \delta_{ij}; \quad
[S^{+}_{i},S^{-}_{i}]_{+} = 2S(S + 1) - 2(S^{z}_{i})^{2};  \\
\nonumber [S^{\mp}_{i},S^{z}_{j}]_{-} = \pm
S^{\mp}_{i}\delta_{ij}; \quad S^{z}_{i} = S(S + 1) -
(S^{z}_{i})^{2} - S^{-}_{i}S^{+}_{i}; \\ \nonumber
(S^{+}_{i})^{2S+1} = 0, \quad (S^{-}_{i})^{2S+1} = 0 \nonumber
\end{eqnarray}
We  omit the term of interaction of the spin with an external
magnetic field for the brevity of notation. The statistical
mechanical problem involving this Hamiltonian was not  exactly
solved, but many approximate
solutions were obtained.\\
To proceed further, it is important to note that for the isotropic
Heisenberg model, the total $z$-component of spin  $S^z_{tot} =
\sum_{i}S^z_{i}$ is a constant of motion, i.e. \begin{equation}
\label{eq.3} [H,S^z_{tot}] = 0
\end{equation}
  There are cases when the total spin is not a constant of motion,
as, for instance, for the Heisenberg model with the dipole terms
added.\\Let us define the eigenstate $|\psi_{0}>$  so that
$S^+_{i}|\psi_{0}> = 0$ for all lattice sites $R_{i}$. It is
clear that $|\psi_{0}>$ is a state in which all the spins are
fully aligned and for which $S^z_{i}|\psi_{0}> = S|\psi_{0}>$. We
also have $$ J_{\vec k} = \sum_{i}e^{(i\vec k \vec R_i)} J(i) =
J_{-\vec k}$$, where the reciprocal vectors $\vec k$ are defined
by cyclic boundary conditions. Then we obtain $$ H |\psi_{0}> = -
\sum_{ij} J(i-j)S^2 = - NS^2 J_{0}$$  Here $N$ is the total number
of ions in the crystal. So, for the isotropic Heisenberg
ferromagnet, the ground state $|\psi_{0}>$ has an energy $-NS^2
J_{0}$.\\ The state $|\psi_{0}>$ corresponds to a total spin
$NS$. \\ Let us consider now the first excited state. This state
can be constructed by creating one unit of spin deviation in the
system. As a result, the total spin is $NS - 1$. The state
$$|\psi_{k}> = \frac {1}{\sqrt {(2SN)}}\sum_{j}e^{(i\vec k \vec R_j)}S^-_{j}|\psi_{0}>$$ is
an eigenstate of $H$ which corresponds to a single magnon of the
energy
\begin{equation}
\label{eq.4} E  (q) = 2S (J_{0} - J_{q})
\end{equation}
Note that the role of translational symmetry, i.e. the regular
lattice of spins, is essential, since the state $|\psi_{k}>$ is
constructed from the fully aligned state by decreasing the spin
at each site and summing over all spins with the phase factor
$e^{i\vec k \vec R_j}$ (we consider the 3-dimensional case only).
It is easy to verify that
$$<\psi_{k}|S^z_{tot} |\psi_{k}> = NS - 1$$ \\ The above
consideration was possible because we knew the exact ground state
of the Hamiltonian . There are many models where this is not the
case. For example, we do not know the exact ground state of a
Heisenberg ferromagnet with dipolar forces and the ground state
of the Heisenberg antiferromagnet.
\subsection{  Itinerant Electron Model } E.Stoner has
proposed an alternative, phenomenological band model of magnetism
of the transition metals in which the bands for electrons of
different spins are shifted in energy in a way that is favourable
to ferromagnetism. The band shift effect is a consequence of
strong intra-atomic correlations. The itinerant-electron picture
is the alternative conceptual picture for
magnetism\cite{her},\cite{mor}. It must be noted that the problem
of antiferromagnetism is a much more complicated
subject\cite{kuze}. The antiferromagnetic state is characterized
by a spatially changing component of magnetization which varies
in such a way that the net magnetization of the system is zero.
The concept of antiferromagnetism of localized spins,  which is
based on the Heisenberg model and two-sublattice Neel ground
state, is relatively well founded contrary to the
antiferromagnetism of delocalized or itinerant electrons .  In
relation to the duality of localized and itinerant electronic
states, G.Wannier showed the importance of the description of the
electronic states which reconcile the band and local (cell)
concept as a matter of principle.
\subsection{  Hubbard Model } There are big difficulties in the
description of the complicated problem of magnetism in a metal
with the $d$ band electrons which  are really neither "local" nor
"itinerant" in a full sense. The Hubbard model\cite{kuz7} is in a
certain sense an intermediate model (the narrow-band model) and
takes into account the specific features of transition metals and
their compounds by assuming that the $d$ electrons form a band,
but are subject to a strong Coulomb repulsion  at one lattice
site.  The Hubbard   Hamiltonian is of the
form\cite{kuze1},\cite{kuze2}
\begin{equation}
\label{eq.5} H =
\sum_{ij\sigma}t_{ij}a^{\dagger}_{i\sigma}a_{j\sigma} +
U/2\sum_{i\sigma}n_{i\sigma}n_{i-\sigma}
\end{equation}
It includes the intra-atomic Coulomb repulsion $U$ and the
one-electron hopping energy $t_{ij}$. The electron correlation
forces electrons to localize in the atomic orbitals which are
modelled here by a complete and orthogonal set of the Wannier wave
functions $[\phi({\vec r} -{\vec R_{j}})]$. On the other hand,
the kinetic energy is reduced when electrons are delocalized. The
band energy of Bloch electrons $\epsilon_{\vec k}$ is defined as
follows: \begin{equation} \label{eq.6}
  t_{ij} = N^{-1}\sum_{\vec k}\epsilon^{d}_{
k} \exp[i{\vec k}({\vec R_{i}} -{\vec R_{j}}] \end{equation}
where $N$ is the number of lattice sites. This conceptually simple
model is mathematically very
complicated\cite{kuze1},\cite{kuze2}.  The Pauli exclusion
principle which does not allow two electrons of common spin to be
at the same site, plays a crucial role. It can be shown, that
under transformation $R H R^{+}$, where $R$ is the spin rotation
operator \begin{equation} \label{eq.7}R = \bigotimes_{j}\exp ({1
\over 2} i \phi \vec \sigma_{j} \vec n)
\end{equation}
the Hubbard Hamiltonian is invariant under spin rotation, {\em
i.e.,} $ R H R^{+} = H$. Here $\phi$ is the angle of rotation
around the unitary axis $\vec n$ and $\vec \sigma$ is the Pauli
spin vector; symbol $\bigotimes_{j}$ indicates a tensor product
over all site subspaces. The summation over $j$ extends to all
sites.\\ The equivalent expression for the Hubbard model that
manifests the property of   rotational invariance explicitly can
be obtained with the aid of the transformation
\begin{equation}
\label{eq.8} \vec S_{i} = {1 \over 2} \sum_{\sigma \sigma'}
a^{\dagger}_{i\sigma} \vec \sigma_{\sigma \sigma'} a_{j\sigma'}
\end{equation}
Then the second term in (\ref{eq.5}) takes the following form
$$n_{i \uparrow}n_{i \downarrow} =
\frac{n_{i}}{2} - \frac{2}{3} \vec S_{i}^{2}$$
As a result we get
\begin{equation}
\label{eq.9} H =
\sum_{ij\sigma}t_{ij}a^{\dagger}_{i\sigma}a_{j\sigma} + U \sum_{i
} (\frac{n_{i}^{2}}{4} - \frac{1}{3} \vec S_{i}^{2})
\end{equation}
The total $z$-component $S^z_{tot}$ commutes with Hubbard
Hamiltonian and the relation (\ref{eq.3}) is valid.
\subsection{Multi-Band Models. Model with $s-d$ Hybridization}
The Hubbard model is the single-band model. It is necessary, in
principle, to take into account the multi-band structure, orbital
degeneracy, interatomic effects and electron-phonon interaction.
The band structure calculations and the experimental studies
showed that for noble, transition and rare-earth metals the
multi-band effects are essential. An important generalization of
the single-band Hubbard model is the so-called model with $s-d$
hybridization\cite{smit},\cite{kiso}. For transition $d$ metals,
investigation of the energy band structure reveals that $s-d$
hybridization processes play an important part. Thus, among the
other generalizations of the Hubbard model that correspond more
closely to the real situation in transition metals, the model
with $s-d$ hybridization serves as an important tool for
analyzing of the multi-band effects. The system is described by a
narrow $d$-like band, a broad $s$-like band and a $s-d$ mixing
term coupling the two former terms. The model Hamiltonian reads
\begin{equation}
\label{eq.10}
   H = H_{d} + H_{s} + H_{s-d }
\end{equation}
The Hamiltonian $H_{d}$ of tight-binding electrons is the Hubbard
model (\ref{eq.5}).
\begin{equation}
\label{eq.11} H_{s} =
\sum_{k\sigma}\epsilon^{s}_{k}c^{\dagger}_{k\sigma}c_{k\sigma}
\end{equation}
is the Hamiltonian of a broad s-like band of electrons.
\begin{equation}
\label{eq.12} H_{s-d} = \sum_{k\sigma} V_{k} (
c^{\dagger}_{k\sigma}a_{k\sigma} +
 a^{\dagger}_{k\sigma}c_{k\sigma})
\end{equation}
is the interaction term which represents a mixture of the
$d$-band and $s$-band electrons. The model Hamiltonian
(\ref{eq.10}) can be interpreted also in terms of a series of
Anderson impurities placed regularly in each site (the so-called
periodic Anderson model ). The model (\ref{eq.10}) is rotationally
invariant also.
\subsection{ Spin-Fermion Model } Many magnetic and electronic
properties of rare-earth metals and compounds ({\it e.g.,}
magnetic semiconductors) can be interpreted in terms of a combined
spin-fermion model~\cite{coq},\cite{kuz4}  that includes the
interacting localized spin and itinerant charge subsystems. The
concept of the $s(d)-f$ model plays an important role in the
quantum theory of magnetism, especially the generalized $d-f$
model, which describes the localized $4f(5f)$-spins interacting
with $d$-like tight-binding itinerant electrons and takes into
consideration the electron-electron interaction.  The total
Hamiltonian of the model is given by
\begin{equation}
\label{eq.13}
   H = H_{d} + H_{d-f}
\end{equation} The Hamiltonian $H_{d}$ of tight-binding electrons is the
Hubbard model (\ref{eq.5}).
  The term $H_{d-f}$ describes the
interaction of the total $4f(5f)$-spins with the spin density of
the itinerant electrons
\begin{equation} \label{eq.14}
 H_{d-f} = \sum_{i}J{\vec \sigma_{i}}{\vec S_{i}}
 = - J
N^{-1/2}\sum_{kq}\sum_{\sigma}[S^{-\sigma}_{-q}a^{\dagger
}_{k\sigma} a_{k+q-\sigma} + z_{\sigma}S^{z}_{-q}a^{\dagger
}_{k\sigma}a_{k+q\sigma}]
\end{equation}
where sign factor $z_{\sigma}$ is given by $$z_{\sigma} = (+ ,
-)\quad - \quad \sigma =  (\uparrow  ,  \downarrow)$$ and
$$S^{-\sigma}_{-q} = \cases {S^{-}_{-q} &- $\sigma = +$ \cr
S^{+}_{-q} &- $\sigma = -$ \cr}$$ In general the indirect
exchange integral $J$ strongly depends on the wave vectors
$J(\vec k; \vec k+ \vec q)$ having its maximum value at $k=q=0$.
We omit this dependence for the sake of brevity  of notation. To
describe the magnetic semiconductors the Heisenberg interaction
term  (\ref{eq.1}) should be added\cite{kuz5},\cite{kuz6} ( the
resulting model is called the modified Zener model ).\\
These   model Hamiltonians (and their simple modifications and
combinations) are the most commonly used models in quantum theory
of magnetism. In our previous paper\cite{kuz2}, where the
detailed analysis of the neutron scattering experiments on
magnetic transition metals and their alloys and compounds was
made, it was concluded that at the level of low-energy
hydrodynamic excitations one cannot distinguish between the
models. The reason for that is the spin-rotation symmetry. In
terms of refs.\cite{pin1},\cite{and2}, the spin waves (
collective waves of the order parameter ) are in a quantum
protectorate precisely in this sense. I will argue below the
latter statement more explicitly.
\section{ Symmetry and Physics of Magnetism }
In many-body interacting systems, the symmetry is important in
classifying different phases and   understanding the phase
transitions between them\cite{ander},\cite{call} . To implement
the QP idea it is necessary to establish the symmetry properties
and corresponding conservation laws of the microscopic models of
magnetism.  The Goldstone theorem states that, in a system with
broken continuous symmetry ( {\it i.e.,} a system such that the
ground state is not invariant under the operations of a
continuous unitary group whose generators commute with the
Hamiltonian ), there exists a collective mode with frequency
vanishing as the momentum goes to zero. For many-particle systems
on a lattice, this statement needs a proper adaptation. In the
above form, the Goldstone theorem is true only if the condensed
and  normal phases have the same translational properties. When
translational symmetry is also broken, the Goldstone mode appears
at   zero frequency but at nonzero momentum, {\it e.g.}, a
crystal and a helical spin-density-wave (SDW) ordering. As has
been noted, this present paper is an attempt to explain the
physical implications involved in the concept of QP for   quantum
theory of magnetism. All the three models considered above, the
Heisenberg, the Hubbard, and the spin-fermion model, are spin
rotationally invariant, $ R H R^{+} = H$. The spontaneous
magnetization of the spin or fermion system on a lattice that
possesses  the spin rotational invariance, indicate on a broken
symmetry  effect, {\it i.e.,} that the physical ground state is
not an eigenstate of the time-independent generators of symmetry
transformations on the original Hamiltonian of the system. As a
consequence, there must exist an excitation mode, that is an
analog of the Goldstone mode for the continuous case (referred to
as "massless" particles). It was shown that both the models, the
Heisenberg model and the band or itinerant electron model of a
solid, are capable of describing the theory of spin waves for
ferromagnetic insulators and metals\cite{kuz2}. In their
paper\cite{kit}, Herring and Kittel showed that in simple
approximations the spin waves can be described equally well in
the framework of the model of localized spins or the model of
itinerant electrons. Therefore the study of, for example, the
temperature dependence of the average moment in magnetic
transition metals in the framework of low-temperature spin-wave
theory does not, as a rule, give any indications in favor of a
particular model.  Moreover, the itinerant electron model (as
well as the localized spin model) is capable of accounting for the
exchange stiffness determining the properties of the transition
region, known as the Bloch wall, which separates adjacent
ferromagnetic domains with different directions of magnetization.
The spin-wave stiffness constant $D$ is defined so that the
energy of a spin wave with a small wave vector $\vec q$ is $E
\sim Dq^2$. To characterize the dynamic behaviour of the magnetic
systems in terms of the quantum many-body theory, the generalized
spin susceptibility (GSS) is a very useful tool\cite{low}. The
GSS is defined by
\begin{equation}
\label{eq.15} \chi (\vec q, \omega) = \int dt << S^{-}_{q}(t),
S^{+}_{-q}>> \exp{(-i\omega t)}
\end{equation}
For the Hubbard model $ S^{-}_{i} =
a^{+}_{i\downarrow}a_{i\uparrow}$. This GSS satisfies the
important sum rule
\begin{equation}
\label{eq.16} \int Im \chi (\vec q, \omega)d\omega =  \pi (
n_{\downarrow} - n_{\uparrow}) = - 2\pi <S^{z}>
\end{equation}
It is possible to check that\cite{kuz2}
\begin{equation}
\label{eq.17} \chi (\vec q, \omega) = -\frac{2<S^{z}>}{\omega} +
\frac{q^2}{\omega^{2}} \{ \Psi(\vec q, \omega) - \frac{1}{q}
\langle [ Q^{-}_{q}, S^{+}_{-q}] \rangle \}
\end{equation}
Here the following notation was used for $q Q^{-}_{q} = [
S^{-}_{q}, H ] $ and $\Psi(\vec q, \omega) = << Q^{-}_{q} |
Q^{+}_{-q}>>_{\omega}$. It is clear from (\ref{eq.17}) that  for
$q = 0$ the GSS (\ref{eq.15}) contains only the first term
corresponding to the spin-wave pole for $q = 0$ which exhausts
the sum rule (\ref{eq.16}). For small $q$, due to the continuation
principle, the GSS $\chi (\vec q, \omega)$ must be dominated by
the spin wave pole with the energy
\begin{equation}
\label{eq.18} \omega = Dq^2 = \frac{1}{2<S^{z}>}\{ q  \langle[
Q^{-}_{q}, S^{+}_{-q}]\rangle - q^{2} \lim_{\omega \rightarrow 0}
\lim_{q \rightarrow 0}
\Psi(\vec q, \omega) \}
\end{equation}
This result is the direct consequence of the spin rotational
invariance and is valid for all the three models considered
above.
\section{Spin Quasiparticle Dynamics }
In this Section, to make the discussion more concrete and to
illustrate the nature of spin excitations in the above described
models, let us consider the generalized spin susceptibility
(GSS), which measures the response of "magnetic" degrees of
freedom to an external perturbation\cite{low}. The GSS is
expressed in terms of the double-time thermal GF of spin variables
~\cite{tyab}\cite{kuz7}, that is defined as
\begin{eqnarray}
\label{eq.19} \chi (q;t - t') = <<S^{+}_{q}(t),S^{-}_{-q}(t')>> =
-i\theta(t - t')<[S^{+}_{q}(t),S^{-}_{-q}(t')]_{-}> = \nonumber\\
1/2\pi \int_{-\infty}^{+\infty} d\omega \exp(-i\omega t) \chi
(q,\omega)
\end{eqnarray} The poles of the GSS determine the energy spectra
of the excitations in the system. The explicit expressions for
the poles are strongly dependent on the model used for the system
and the character of
approximations\cite{kuz2},\cite{low}.\\
The next step in description of the spin quasiparticle dynamics is
to write down the equation of motion for the GF. Our attention is
focused on the spin dynamics of the models. To describe
self-consistently the spin dynamics of the models one should take
into account the full algebra of relevant operators of the
suitable "spin modes", which are appropriate for the case.
\subsection{Spin Dynamics of the Hubbard Model}
Theoretical calculations of the GSS in transition $3d$ metals
have been largely based on the single-band Hubbard
Hamiltonian\cite{low}. The GSS for this case reads
\begin{equation}
\label{eq.20} \chi (q,\omega) =  <<\sigma^{+}_{q}\vert
\sigma^{-}_{-q} >>_{\omega}
\end{equation}
Here $$\sigma^{+}_{k} = \sum_{p}
a^{\dagger}_{k\uparrow}a_{k+p\downarrow} ;\quad \sigma^{-}_{k} =
\sum_{p} a^{\dagger}_{k\downarrow}a_{k+p\uparrow} $$ The result
of the RPA calculation\cite{low} has the following form
\begin{equation}
\label{eq.21}  \chi (q,\omega) = <<\sigma^{+}_{q}\vert
\sigma^{-}_{-q} >>_{\omega} = \frac {\chi^{0} (q,\omega)}{1 - U
\chi^{0} (q,\omega) }
\end{equation}
where
\begin{equation}
\label{eq.22}  \chi^{0} (q,\omega) = N^{-1} \sum_{k} \frac
{n_{k\uparrow} - n_{k+q\downarrow}}{\omega + \epsilon^{d}_{k+q} -
\epsilon^{d}_{k} - \Delta }
\end{equation}
\begin{equation}
\label{eq.23} \Delta = {U \over N} \sum_{k} (n_{k\downarrow} -
n_{k \uparrow})
\end{equation}
The excitation spectrum of the Hubbard model determined by the
poles of susceptibility (\ref{eq.22})  is shown schematically in
fig.1. The experimental data for three typical magnetic material
are listed in Table 1. Note, that typically $q_{max} \le 0.75
k_{F}$.
\begin{table}
\begin{center}
\caption{  EXPERIMENTAL DATA for TRANSITION  METALS}
\end{center}
\begin{center}
\begin{tabular}{|l|c|c|c|c|c|c| } \hline
Element $\backslash$ $Data$ & $ T_{c}$ & $D \quad meV A^2$ & $\mu \quad \mu_{B}$ & $\Delta \quad eV$ & $q_{max}$   \\
 \hline
$Fe$ & $1043$ K &$ 280$ & $2.177  $  &- & -  \\
 \hline
$Co$ & $1403$ K & 510 & $1.707  $  & $0.91  $  & - \\
 \hline
$Ni$ & $ 631$ K& $433  $ &$ 0.583$ & $0.5 \pm 0.1   $  & $0.8  A^{-1}$ \\
 \hline
$MnSi$ & $30$ K & 52 & $0.4 $  & -  & - \\
 \hline
\end{tabular}
\end{center}
\end{table}
\subsection{Spin Dynamics of the Spin-Fermion Model}
When the goal is to describe self-consistently the quasiparticle
dynamics  of two interacting subsystems the situation is more
complicated. For the spin-fermion model (\ref{eq.14}) the
relevant algebra of operators should be described by the 'spinor'
${\vec S_{i}\choose \vec \sigma_{i}}$ ("relevant degrees of
freedom")\cite{kuz4}. Once this has been done, one should
introduce the generalized matrix spin susceptibility of the form
\begin{equation} \label{eq.24}
\pmatrix{ <<S^{+}_{k}\vert S^{-}_{-k}>> &
<<S^{+}_{k}\vert \sigma^{-}_{-k}>> \cr <<\sigma^{+}_{k}\vert
S^{-}_{-k}>> & <<\sigma^{+}_{k}\vert \sigma^{-}_{-k}>> \cr} = \hat
\chi(k,\omega) \end{equation} The spectrum of quasiparticle
excitations without damping
follows from the poles of the generalized mean-field susceptibility.\\
Let us write down explicitly the first matrix element
$\chi^{11}_{0}$
\begin{equation} \label{eq.25}
<<S^{+}_{q} \vert S^{-}_{-q}>>^{0} = \frac
{2JN^{-1/2}<S^{z}_{0}>}{ \omega - JN^{-1}(n_{\uparrow} -
n_{\downarrow}) + 2J^{2}N^{-1/2}<S^{z}_{0}> (1 - U\chi^{df}_{0}
)^{-1}\chi^{df}_{0}}
\end{equation}
where
\begin{eqnarray} \label{eq.26}
\chi ^{df}_{0}(k,\omega) = N^{-1} \sum_{p} \frac {
(n_{p+k\downarrow} - n_{p\uparrow})}{\omega_{p,k}}\\
\omega_{p,k} = (\omega + \epsilon^{d}_{p}   - \epsilon^{d}_{p+k} - \Delta) \\
\nonumber \Delta =
2JN^{-1/2}<S^{z}_{0}> - UN^{-1}(n_{\uparrow} - n_{\downarrow}) \\
\nonumber
\end{eqnarray}
 This result can be considered  as reasonable approximation
for description of the dynamics of localized spins in heavy
rare-earth
metals like $Gd$.  (c.f.~\cite{coq} ).\\
The magnetic excitation spectrum that follows from the GF
(\ref{eq.24}) consists of three branches - the acoustic  spin
wave, the optic spin wave and the Stoner continuum~\cite{kuz4}.
In the hydrodynamic limit, $q \rightarrow 0$, $\omega \rightarrow
0$ the GF  (\ref{eq.24}) can be written as
\begin{equation} \label{eq.27}
<<S^{+}_{q} \vert S^{-}_{-q}>>^{0} = \frac {2N^{-1/2} < \tilde
S^{z}_{0}>}{ \omega - E(q)}
\end{equation}
where the acoustic  spin wave energies are given by
\begin{equation} \label{eq.28}
E(q) = Dq^{2} = \frac {1/2 \sum_{k}(n_{k\uparrow} +
n_{k\downarrow})(\vec q \frac {\partial}{\partial \vec k})^{2}
\epsilon^{d}_{\vec k} + (2\Delta)^{-1}\sum_{k}(n_{k\uparrow} -
n_{k\downarrow})(\vec q \frac {\partial}{\partial \vec k}
\epsilon^{d}_{\vec k})^{2}}{ 2N^{1/2}<S^{z}_{0}> + (n_{\uparrow} -
n_{\downarrow})}
\end{equation}
and
\begin{equation} \label{eq.30}
< \tilde S^{z}_{0}> = <S^{z}_{0}> [ 1 + \frac {(n_{\uparrow} -
n_{\downarrow})}{ 2N^{3/2} <S^{z}_{0}>}]^{-1}
\end{equation}
In GMF approximation the density of itinerant electrons ( and the
band splitting $\Delta$) can be evaluated by solving the equation
\begin{equation} \label{eq.31}
n_{\sigma} = \sum_{k}<a^{+}_{k\sigma}a_{k\sigma}> = \sum_{k} [\exp
(\beta(\epsilon^{d}_{k} + UN^{-1}n_{-\sigma} -
JN^{-1/2}<S^{z}_{0}> - \epsilon_{F})) + 1]^{-1}
\end{equation}
Hence, the stiffness constant $D$ can be expressed by the
parameters of the Hamiltonian  (\ref{eq.13}).\\ The spectrum of
the Stoner excitations is given by~\cite{kuz4}
\begin{equation} \label{eq.32}
E^{St}(q) = \epsilon^{d}_{k+q} - \epsilon^{d}_{k} + \Delta
\end{equation}
If we consider the optical spin wave branch then by direct
calculation one can easily show that
\begin{eqnarray} \label{eq.33}
E_{opt}(q) = E^{0}_{opt} + D(UE_{opt}/J\Delta - 1)q^{2} \nonumber \\
E^{0}_{opt} = J(n_{\uparrow} - n_{\downarrow}) + 2J<S^{z}_{0}>
\end{eqnarray}
From the equation (33) one also finds the GF of itinerant spin
density in the generalized mean field approximation
\begin{equation} \label{eq.34}
<<\sigma^{+}_{k}\vert \sigma^{-}_{-k}>>^{0}_{\omega}  = \frac
{\chi^{df}_{0}(k,\omega)}{1 - [ U - \frac {
2J^{2}<S^{z}_{0}>}{\omega - J( n_{\uparrow} - n_{\downarrow})}]
\chi^{df}_{0}(k,\omega)} \end{equation}
\subsection{Spin Dynamics of the Multi-Band Model}
Now let us calculate the GSS for the Hamiltonian (\ref{eq.10}).
In general, one should introduce the generalized matrix spin
susceptibility of the form
\begin{equation} \label{eq.35}
\pmatrix{ <<\sigma^{-}_{q}\vert \sigma^{+}_{-q}>> &
<<\sigma^{-}_{q}\vert s^{-}_{-q}>> \cr <<s^{+}_{q}\vert
\sigma^{-}_{-q}>> & <<s^{-}_{q}\vert s^{+}_{-q}>> \cr} = \hat
\chi(q,\omega) \end{equation} Here $$s^{+}_{k} = \sum_{q}
c^{\dagger}_{k\uparrow}c_{k+q\downarrow} ;\quad s^{-}_{k} =
\sum_{q} c^{\dagger}_{k\downarrow}c_{k+q\uparrow} $$ Let us
consider for brevity the calculation of the Green function
$<<\sigma^{-}_{q}\vert \sigma^{+}_{-q}>>$.  According to
ref.\cite{low}, the object now is to calculate the Green function
$<<\theta_{k}(q) = a^{\dagger}_{k+q\downarrow}a_{k\uparrow} \vert
\sigma^{+}_{-q}>>_{\omega}$. In the random phase approximation
(RPA), the equations of motion for the relevant Green functions
are reduced to the closed form
\begin{eqnarray}\label{eq.36}
(\omega + \epsilon^{d}_{\uparrow}(k+q) -
\epsilon^{d}_{\downarrow}(k )) <<\theta_{k}(q) \vert
\sigma^{+}_{-q}>>_{\omega} = (n_{k+q\downarrow} - n_{k
\uparrow})A(q,\omega) \\ \nonumber - V_{k+q} <<
c^{\dagger}_{k+q\downarrow} a_{k\uparrow} \vert
\sigma^{+}_{-q}>>_{\omega} + V_{k} <<a^{\dagger}_{k+q\downarrow}
a_{k\uparrow} \vert \sigma^{-}_{-q}>>_{\omega}\\ \label{eq.37}
(\omega - \epsilon^{d}_{\downarrow}(k) + \epsilon^{s}_{k+q})
<<c^{\dagger}_{k+q\downarrow} a_{k\uparrow} \vert
\sigma^{+}_{-q}>>_{\omega} =  < c^{\dagger}_{k+q\downarrow}
a_{k+q\downarrow} > A(q,\omega) \\ \nonumber + V_{k} <<
c^{\dagger}_{k+q\downarrow} c_{k\uparrow} \vert
\sigma^{+}_{-q}>>_{\omega} - V_{k+q} << \theta_{k}(q)  \vert
\sigma^{-}_{-q}>>_{\omega}\\ \label{eq.38} (\omega -
\epsilon^{d}_{\uparrow}(k+q) - \epsilon^{s}_{k})
<<a^{\dagger}_{k+q\downarrow} c_{k\uparrow} \vert
\sigma^{+}_{-q}>>_{\omega} =  < a^{\dagger}_{k \uparrow} c_{k
\uparrow} > A(q,\omega) \\ \nonumber + V_{k} <<
a^{\dagger}_{k+q\downarrow} a_{k\uparrow} \vert
\sigma^{+}_{-q}>>_{\omega} - V_{k+q} <<
c^{\dagger}_{k+q\downarrow} c_{k\uparrow} \vert
\sigma^{+}_{-q}>>_{\omega}\\ \label{eq.39} (\omega +
\epsilon^{s}_{k+q}  - \epsilon^{s}_{k})
<<c^{\dagger}_{k+q\downarrow} c_{k\uparrow} \vert
\sigma^{+}_{-q}>>_{\omega} =   \\ \nonumber + V_{k} <<
c^{\dagger}_{k+q\downarrow} a_{k\uparrow} \vert
\sigma^{+}_{-q}>>_{\omega} - V_{k+q} <<
a^{\dagger}_{k+q\downarrow} c_{k\uparrow} \vert
\sigma^{+}_{-q}>>_{\omega}
\end{eqnarray}
Here the following definitions were introduced
\begin{eqnarray} \label{eq.40}
\epsilon^{d}_{\sigma}(k ) = \epsilon^{d}_{k} + {U \over N}
\sum_{p}<a^{\dagger}_{p\sigma}a_{p\sigma}> \\ \nonumber A(q,
\omega) = 1 - {U \over N} <<\sigma^{-}_{q}\vert
\sigma^{+}_{-q}>>_{\omega}
\end{eqnarray}
To truncate the hierarchy of Green functions equations
(\ref{eq.36}) - (\ref{eq.39}) the RPA linearization was used
\begin{eqnarray}
\label{eq.41} [\theta_{k}(q),H_{d}]_{-} \sim  (\epsilon^{d}_{k} -
\epsilon^{d}_{k+q})\theta_{k}(q) + \Delta \theta_{k}(q) \\
\nonumber- {U \over N} \sum_{p}(<a^{\dagger}_{k+q
\downarrow}a_{k+q\downarrow}>
- <a^{\dagger}_{k  \uparrow}a_{k
\uparrow}>)\theta_{p}(q)\\ \nonumber [a^{\dagger}_{k+q\downarrow}
c_{k\uparrow},H_{d}]_{-} \sim -
\epsilon^{d}_{k+q} a^{\dagger}_{k+q\downarrow}c_{k\uparrow} - \\
\nonumber {U \over N} \sum_{p} n_{p\uparrow}
<a^{\dagger}_{k+q\downarrow}c_{k\uparrow}> + {U \over N}
<a^{\dagger}_{k \uparrow}c_{k\uparrow}> \sum_{p}
a^{\dagger}_{p+q\downarrow}a_{p\uparrow}
\end{eqnarray}
  Now,
we will use these equations to determine the spin susceptibility
of $d$-electron subsystem in the random phase approximation. It
can be shown that
\begin{equation} \label{eq.42}
   \chi (q,\omega) = <<\sigma^{-}_{q}\vert
\sigma^{+}_{-q} >>_{\omega} = \frac {\chi^{MF} (q,\omega)}{1 - U
\chi^{MF} (q,\omega) }
\end{equation}
We introduced here the notation $\chi^{MF} (q,\omega) $ for the
mean field susceptibility to distinguish it from the
$\chi^{0}(q,\omega) $ (\ref{eq.22}).\\
The expression for the $\chi^{MF} (q,\omega) $  is of the form
\begin{eqnarray}
\label{eq.43} \chi^{MF} (q,\omega) = {1 \over N} \sum_{k} \{ (
n_{k+q\downarrow} - n_{k\uparrow}) [ - \vert V_{k} \vert^{2}
\Bigl( ( \omega + \epsilon^{d}_{\downarrow}(k) +
\epsilon^{s}_{k+q}) \\ \nonumber + ( \omega +
\epsilon^{d}_{\uparrow}(k+q) - \epsilon^{s}_{k}) \Bigr ) \\
\nonumber + ( \omega + \epsilon^{s}_{k+q} - \epsilon^{s}_{k})(
\omega + \epsilon^{d}_{\uparrow}(k+q) - \epsilon^{s}_{k}) ( \omega
- \epsilon^{d}_{\downarrow}(k) + \epsilon^{s}_{k+q})] \\
\nonumber - ( \omega + \epsilon^{s}_{k+q} -
\epsilon^{s}_{k})[V_{k}<a^{\dagger}_{k \uparrow}c_{k\uparrow}>(
\omega - \epsilon^{d}_{\downarrow}(k) - \epsilon^{s}_{k+q}) + \\
\nonumber V_{k}<c^{\dagger}_{k+q\downarrow}a_{k+q\downarrow}>(
\omega + \epsilon^{d}_{\uparrow}(k+q) - \epsilon^{s}_{k})] \}
R^{-1}
\end{eqnarray}
where
\begin{eqnarray}
\label{eq.44} R  = \{ - \vert V_{k} \vert^{2} \Bigl( ( \omega +
\epsilon^{d}_{\uparrow}(k+q) - \epsilon^{d}_{\downarrow}(k))(
\omega + \epsilon^{d}_{\uparrow}(k+q) - \epsilon^{s}_{k}) \\
\nonumber + ( \omega - \epsilon^{d}_{\downarrow}(k) +
\epsilon^{s}_{k+q})( \omega + \epsilon^{s}_{k+q} -
\epsilon^{s}_{k}) + ( \omega + \epsilon^{d}_{\uparrow}(k+q) -
\epsilon^{d}_{\downarrow}(k))( \omega -
\epsilon^{d}_{\downarrow}(k) + \epsilon^{s}_{k+q}) \\ \nonumber +
( \omega + \epsilon^{d}_{\uparrow}(k+q) - \epsilon^{s}_{k})(
\omega + \epsilon^{s}_{k+q} - \epsilon^{s}_{k}) \Bigr ) \\
\nonumber + ( \omega + \epsilon^{d}_{\uparrow}(k+q) -
\epsilon^{d}_{\downarrow}(k))( \omega -
\epsilon^{d}_{\downarrow}(k) + \epsilon^{s}_{k+q})( \omega +
\epsilon^{d}_{\uparrow}(k+q) - \epsilon^{s}_{k})( \omega +
\epsilon^{s}_{k+q} - \epsilon^{s}_{k}) \}
\end{eqnarray}
Note, that if $V_{k} = 0$ then, $\chi^{MF} (q,\omega) $ is
reduced precisely to $\chi^{0} (q,\omega) $ (\ref{eq.22}).\\
The spectrum of quasiparticle excitations corresponds   to the
poles of the spin susceptibility (\ref{eq.22}); it corresponds to
the spin-wave modes and to the Stoner-like spin-flip modes. Let
us discuss first the question about the existence of a spin-wave
pole among the set of poles of the susceptibility (\ref{eq.42}).
If we set $q = 0$ in (\ref{eq.43}) the secular equation for poles
becomes
\begin{eqnarray}
\label{eq.45} 1  =  {U \over N} \sum_{k} \{ ( n_{k \downarrow} -
n_{k\uparrow}) [ - \vert V_{k} \vert^{2}  ( 2 \omega - \Delta )
\\ \nonumber + \omega ( \omega + \epsilon^{d}_{\uparrow}(k ) -
\epsilon^{s}_{k}) ( \omega - \epsilon^{d}_{\downarrow}(k) +
\epsilon^{s}_{k })] \\ \nonumber - \omega [ V_{k}<a^{\dagger}_{k
\uparrow}c_{k\uparrow}> ( \omega - \epsilon^{d}_{\downarrow}(k) +
\epsilon^{s}_{k }) + V_{k}<c^{\dagger}_{k \downarrow}a_{k
\downarrow}>( \omega + \epsilon^{d}_{\uparrow}(k ) -
\epsilon^{s}_{k})] \}  \\ \nonumber \Bigl (
 - \vert V_{k} \vert^{2}  ( 2 \omega + \Delta )^{2} + \\ \nonumber \omega
( \omega - \epsilon^{d}_{\downarrow}(k) + \epsilon^{s}_{k})(
\omega + \epsilon^{d}_{\uparrow}(k) - \epsilon^{s}_{k})( \omega +
\epsilon^{s}_{k} - \epsilon^{s}_{k}) \Bigr )^{-1}
\end{eqnarray}
which is satisfied if $\omega = 0$. It follows from general
considerations of Section 4 that when the wave length of a spin
wave is very long (hydrodynamic limit ), its energy $E(q)$ must
be related to the wave number $q$ by $ E(q) = Dq^{2}$. Thus the
solution for the equation
\begin{equation}
\label{eq.46}
 1 = U \chi^{MF} (q,\omega)
\end{equation}
exists which has the property $\lim_{q \rightarrow 0} E(q) = 0 $
and this solution corresponds to a spin-wave excitation in the
multiband model with $s-d$ hybridization (\ref{eq.42}). Thus we
derived a formula (\ref{eq.42}) for the dynamic spin
susceptibility $\chi (q,\omega)$ in RPA and shown, that it can be
calculated in terms of the mean field spin susceptibility
$\chi^{MF} (q,\omega)$ by analogy with the single-band
Hubbard model.\\
Let us consider the poles of the $ \chi^{MF} (q,\omega)$. It is
instructive to remark that the Hamiltonian  (\ref{eq.10}) can be
rewritten in the mean field approximation as
\begin{equation}
\label{eq.47} H^{MF} = \sum_{k\sigma}\epsilon^{d}_{\sigma}(k)
a^{\dagger}_{k\sigma}a_{k\sigma} + \sum_{k\sigma}\epsilon^{s}_{k}
c^{\dagger}_{k\sigma}c_{k\sigma} + \sum_{k\sigma} V_{k} (
c^{\dagger}_{k\sigma}a_{k\sigma} +
 a^{\dagger}_{k\sigma}c_{k\sigma})
\end{equation}
The Hamiltonian (\ref{eq.46}) can be diagonalized by the
Bogoliubov $(u, v)$-transformation
\begin{equation}
\label{eq.48}   a_{k\sigma} = u_{k \sigma}  \alpha_{k \sigma}  +
v_{k \sigma} \beta_{k \sigma}; \quad  c_{k \sigma} = u_{k \sigma}
\beta_{k \sigma} - v_{k \sigma} \alpha_{k \sigma}
\end{equation}
The result of diagonalization is
\begin{equation}
\label{eq.49} H^{MF} = \sum_{k\sigma} (\omega_{1k\sigma}
\alpha^{\dagger}_{k\sigma}\alpha_{k\sigma} + \omega_{2k\sigma}
 \beta^{\dagger}_{k\sigma}\beta_{k\sigma})
\end{equation}
where
\begin{eqnarray}
\label{eq.50} \omega{1\atop 2 k\sigma}  = \\ \nonumber 1/2[(
\epsilon^{d}_{\sigma}(k) + \epsilon^{s}_{k})  \pm \sqrt { (
\epsilon^{d}_{\sigma}(k) - \epsilon^{s}_{k})^{2} +  4 \vert V_{k}
\vert^{2}}]
\end{eqnarray}
\begin{equation}
\label{eq.51} {u^{2}_{k\sigma} \atop v^{2}_{k\sigma} } = \Bigl [ 1
+ \frac{(\omega{1\atop 2 k\sigma} -
\epsilon^{d}_{\sigma}(k))^{2}} {V_{k}^{2}} \Bigr ]^{-1}
\end{equation}
Then we find
\begin{eqnarray}
\label{eq.52} \chi^{MF}  (q,\omega) = \frac{1}{N} \sum_{k} \{
u^{2}_{k+q \downarrow}u^{2}_{k\uparrow} \frac{( n^{\alpha}_{k
\uparrow} - n^{\alpha}_{k+q\downarrow})}{( \omega +  \omega_{1k+q
\downarrow} - \omega_{1k\uparrow} )}  \\ \nonumber + v^{2}_{
k+q\downarrow}v^{2}_{k\uparrow} \frac{( n^{\beta}_{k \uparrow} -
n^{\beta}_{k+q\downarrow})}{( \omega +  \omega_{2k+q \downarrow}
- \omega_{2k\uparrow} )} + u^{2}_{k+q\downarrow}v^{2}_{k\uparrow}
\frac{( n^{\beta}_{k \uparrow} - n^{\alpha}_{k+q\downarrow})}{(
\omega +  \omega_{1k+q \downarrow}
- \omega_{2k\uparrow} )} + \\
\nonumber v^{2}_{k+q\downarrow}u^{2}_{k\uparrow} \frac{(
n^{\alpha}_{k \uparrow} - n^{\beta}_{k+q\downarrow})}{( \omega +
\omega_{2k+q \downarrow} - \omega_{1k\uparrow} )} \}
\end{eqnarray}
The present consideration shows that for the correlated model
with $s-d$ hybridization the spectrum of spin quasiparticle
excitations is modified in comparison with the single-band
Hubbard model.
\section{ Quasiparticle Excitation Spectra and Neutron Scattering}
\begin{figure}[t] 
\includegraphics[width=4in]{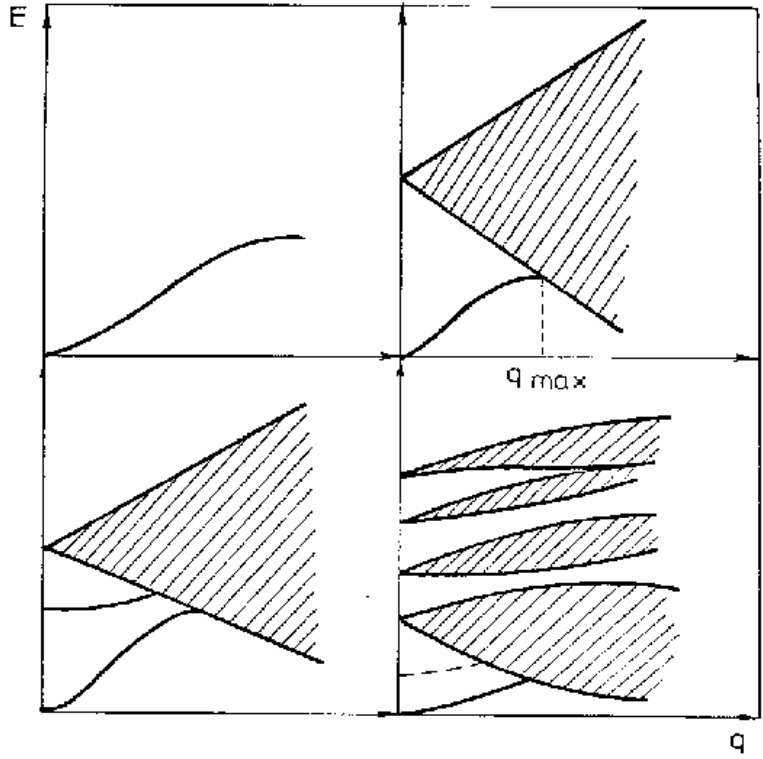}%
 \caption{\label{f.1}  Schematic form of the excitation spectra of the four
microscopic models of magnetism: a) upper-left, the Heisenberg
model; b) upper-right, the Hubbard model; c) down-left, the
modified Zener (spin-fermion) model; d) down-right, the multiband
Hubbard model. }
\end{figure}

The investigation of the spectrum of magnetic excitations of
transition and rare-earth metals and their compounds  is of great
interest for refining our theoretical model representations about
the nature of magnetism. Experiments that probe the
quasi-particle states could shed new light on the fundamental
aspects of the physics of magnetism. The most direct and
convenient method  of experimental study of the spectrum of
magnetic excitations is the method of inelastic scattering of
thermal neutrons . It is known experimentally that the spin wave
scattering of slow neutrons in transition metals and compounds
can be described on the basis of the Heisenberg model. On the
other hand, the mean magnetic moments  of the ions in solids
differ appreciably from the atomic values and are often
fractional.  The main statement of the present consideration is
that the excitation spectrum of the Hubbard model and some of its
modifications is of considerable interest from the point of view
of the choice of the relevant microscopic model. Let us consider
the neutron scattering cross section which  is proportional to
the imaginary part of the GSS\cite{low}
\begin{eqnarray}
\label{eq.53} \frac {d^{2} \sigma}{ d\Omega d\omega} = \Bigl (
\frac{\gamma e^{2}}{m_{e} c^{2}} \Bigr )^{2} \vert F(q) \vert^{2}
(\frac{-1}{2}) \frac{k'}{k} (1 + \tilde q^{2}_{z}) \\ \nonumber  [
( N(\omega ) + 1) Im \chi  ( -\vec q, \omega) + N( -\omega )Im
\chi  (\vec q,  \omega)]
\end{eqnarray}
Here $N(E(k))$ is the Bose distribution function $ N(E(k)) = [
\exp ( E(k)\beta) - 1 ]^{-1}$. To calculate the cross section
(\ref{eq.53}), we obtain from (\ref{eq.42}) the imaginary part of
the susceptibility, namely
\begin{equation}
\label{eq.54}   Im \chi (q,\omega)   =    \frac {Im \chi^{MF}
(q,\omega)}{[1 - U Re \chi^{MF} (q,\omega)]^{2} + [U Im \chi^{MF}
(q,\omega)]^{2} }
\end{equation}
The spin wave pole occurs where $Im \chi^{MF} (q,\omega) $ tends
to zero\cite{low}. In this case, we can in (\ref{eq.54}) take the
limit $Im \chi^{MF} (q,\omega) \rightarrow 0$ so that
\begin{equation}
\label{eq.55}  U Im \chi (q,\omega)  \sim -\pi \delta  [1 - U Re
\chi^{MF} (q,\omega)]
\end{equation}
but
\begin{equation}
\label{eq.56}    1 - U Re \chi^{MF} (q \rightarrow 0,\omega
\rightarrow 0) \sim b^{-1} ( \omega - E(q))
\end{equation}
and thus
\begin{equation}
\label{eq.57}    Im \chi  (q \rightarrow 0,\omega \rightarrow 0)
\sim -\pi {b \over U} \delta ( \omega - E(q))
\end{equation}
Here $b$ is a certain constant, which can be numerically
calculated and $E(q)$ is the acoustic spin wave pole $E(q
\rightarrow 0)  = 0$. \\ Turning now to the calculation of the
cross section (\ref{eq.53}), we obtain the following result
\begin{eqnarray}
\label{eq.58} \frac {d^{2} \sigma}{ d\Omega d\omega} \sim \Bigl (
\frac{\gamma e^{2}}{m_{e} c^{2}} \Bigr )^{2} \vert F(q) \vert^{2}
(\frac{ 1}{4}) \frac{k'}{k} (1 + \tilde q^{2}_{z})N {b \over U}
\\ \nonumber \sum_{p}[ N(
 E(p) ) \delta ( \omega + E(p)) + ( N( E(p))  + 1) \delta ( \omega - E(p))]
\end{eqnarray}
According to   formula (\ref{eq.58}), the cross section for the
acoustic spin wave scattering will be identical for the
Heisenberg and Hubbard (single-band and multiband) model.  So, at
the level of low-energy, hydrodynamic excitations one cannot
distinguish between the models. However, for the Hubbard model,
the poles of the GSS will contain, in addition to acoustic
spin-wave pole, the continuum of the Stoner excitations $
E^{St}(q) = \epsilon_{k+q} - \epsilon_{q} + \Delta$, as is shown
on fig.~\ref{f.1}.
%
%
The spectra of the spin-fermion  model and multiorbital
(multi-band) Hubbard model are shown for   comparison. \\ The
cross section (\ref{eq.58}) does not include the contribution
arising from the scattering by Stoner excitations, i.e. that
determined by $  \chi^{MF} (q,\omega) $. It was shown in
paper\cite{kuz2} that in a single-band Hubbard model of
transition metal in the limit when the wave vector of the
elementary excitations goes to zero, the acoustic spin-wave mode
dominates the inelastic neutron scattering, and the contribution
to the cross section due to Stoner-mode scattering goes to zero.
It was shown that the Stoner-mode scattering intensity does not
become comparable to the spin-wave scattering intensity until $ q
= 0.9 q_{max}$ (see fig.1). Here $q_{max}$ is the value of $q$
when the spin wave enters the continuum. For large values of $q$
and $\omega$ the energy gap $\Delta$ for spin flipping Stoner
excitations may be overcome. In this case
\begin{equation}
\label{eq.59}    Im \chi (q,\omega)  \sim Im \chi^{MF} (q,\omega)
\end{equation}
From (\ref{eq.52}) we obtain for $Im \chi^{MF} (q,\omega) $ the
result
\begin{eqnarray}
\label{eq.60} Im \chi^{MF}  (q,\omega) = \frac{-\pi}{N} \sum_{k}
\{ u^{2}_{k+q \downarrow}u^{2}_{k\uparrow}  ( n^{\alpha}_{k
\uparrow} - n^{\alpha}_{k+q\downarrow}) \delta ( \omega +
\omega_{1k+q \downarrow} - \omega_{1k\uparrow} )   \\ \nonumber +
v^{2}_{ k+q\downarrow}v^{2}_{k\uparrow}  ( n^{\beta}_{k \uparrow}
- n^{\beta}_{k+q\downarrow})\delta( \omega +  \omega_{2k+q
\downarrow} - \omega_{2k\uparrow} ) \\ \nonumber +
u^{2}_{k+q\downarrow}v^{2}_{k\uparrow}  ( n^{\beta}_{k \uparrow}
- n^{\alpha}_{k+q\downarrow})\delta( \omega +  \omega_{1k+q
\downarrow}
- \omega_{2k\uparrow} )  + \\
\nonumber v^{2}_{k+q\downarrow}u^{2}_{k\uparrow}  ( n^{\alpha}_{k
\uparrow} - n^{\beta}_{k+q\downarrow})\delta( \omega +
\omega_{2k+q \downarrow} - \omega_{1k\uparrow} )  \}
\end{eqnarray}
Now it follows from (\ref{eq.60}) that $Im \chi^{MF} (q,\omega) $
is non-zero only for values of the energies equal to the energies
of the Stoner-type excitations
\begin{eqnarray}
\label{eq.61}  E^{St}_{1}(q) = \omega_{1k\uparrow}  - \omega_{1k+q \downarrow}    \\
\nonumber E^{St}_{2}(q) = \omega_{2k\uparrow} - \omega_{2k+q
\downarrow}   \\ \nonumber E^{St}_{3}(q) =
\omega_{2k\uparrow} -  \omega_{1k+q \downarrow}   \\
\nonumber  E^{St}_{4}(q) = \omega_{1k\uparrow} - \omega_{2k+q
\downarrow}
\end{eqnarray}
With (\ref{eq.60}) and (\ref{eq.61}) we obtain
\begin{eqnarray}
\label{eq.62} \frac {d^{2} \sigma}{ d\Omega d\omega} \sim \Bigl (
\frac{\gamma e^{2}}{m_{e} c^{2}} \Bigr )^{2} \vert F(q) \vert^{2}
(\frac{ 1}{4}) \frac{k'}{k} (1 + \tilde q^{2}_{z}) \frac{ N}{\pi}\\
\nonumber  [ ( N(\omega ) + 1) Im \chi^{MF}  ( -\vec q, \omega) +
N( -\omega )Im \chi^{MF}  (\vec q,  \omega)]
\end{eqnarray}
Although for the single-band model the Stoner-mode scattering
cross section remains relatively small until $q$ is fairly close
to  $q_{max}$, it can be shown ( see\cite{kuz2}) that in the
multiband models the Stoner-mode cross section may become
reasonably large for much smaller scattering vector.\\ The
essential result of the present consideration is the calculation
of the GSS for the model with $s-d$ hybridization which is more
realistic for transition metals than the single-band Hubbard
model. The present qualitative treatment shows that   a two-band
picture of inelastic neutron scattering is modified in comparison
with the single-band Hubbard model. We have found that the
long-wave-length acoustic spin-wave  excitations should exist in
this model and that in the limit $(\lim_{\omega \rightarrow 0}
\lim_{q \rightarrow 0} )$, the acoustic spin-wave mode dominates
the inelastic neutron scattering. The spin-wave part of the cross
section is renormalized only quantitatively. The cross section
due to Stoner-mode scattering is qualitatively modified because
of occuring of the four intersecting Stoner-type sub-bands which
may lead to the modification of the spin wave intensity fall off
with increasing energy transfer. The intersection point $q_{max}$
can be essentially renormalized.
\section{ Conclusions  }
In summary, in this article, the logic of an approach to the
quantum theory of magnetism based on the idea of the QP was
described. There is an important aspect of this consideration,
which is seen to be the key principle for the interpretation of
the spin quasiparticle dynamics of the microscopic models of
magnetism.
\\ To
summarize, the usefulness of the QP concept for physics of
magnetism derives from the following features. From our point of
view , the clearest difference between the models is manifested
in the spectrum of magnetic excitations. The model of correlated
itinerant electrons and the spin-fermion model have more
complicated spectra than the model of localized spins (see
fig.~\ref{f.1}) . Since the structure of the GSS and the form of
its poles are determined by the choice of the model Hamiltonian
of the system and the approximations made in its calculation, the
results of neutron scattering experiments can be used to judge
the adequacy of the microscopic models. However, it should be
emphasized that to judge reliably the applicability of a
particular model, it is necessary to measure the susceptibility
(the cross section) at all points of the reciprocal space and for
a wide interval of temperatures, which is not always permitted by
the existing experimental techniques. Thus,   further development
of experimental facilities will provide a base for further
refining of the theoretical models and conceptions about the
nature of magnetism. In terms of ref.\cite{pin1}, to judge which
of the models is more suitable, it is necessary to escape the QP.
This can be done by measurements in the high $(\vec q, \omega )$
region, where $(\vec q \sim q_{max}, E \sim \Delta )$ .\\

The following statements can now be made as to our
analysis and its results. In this paper, we shown that
quasiparticle dynamics of magnetic materials   can be reasonably
understood by using the simplified, but workable models of
interacting spins and electrons on a lattice in the light of the
QP concept. The spectrum of magnetic excitations of the Hubbard
model reflects the dual behaviour of the magnetically active
electrons in transition metals and their compounds. The general
properties of rotational invariance of the model Hamiltonians
show  that the presence of a spin-wave acoustic pole in the
generalized magnetic susceptibility is a direct consequence of
the rotational symmetry of the system. Thus, the acoustic
spin-wave branch reflects  a certain degree of localization of
the relevant electrons; the characteristic quantity $D$, which
determines the spin wave stiffness, can be measured directly in
neutron experiments. In contrast, in the simplified Stoner model
of band ferromagnetism the acoustic spin-waves do  not exist.
There is  a continuum of single-particle Stoner excitations only.
The presence of the Stoner continuum for the spectrum of
excitations of the Hubbard model is a manifestation of the
delocalization of the magnetic electrons.  Since the Stoner
excitations do not arise in the Heisenberg model, their direct
detection and detailed investigation by means of neutron
scattering is one of the most intriguing problems of the
fundamental physics of magnetic state. Concerning the QP notion
studied in the present paper, an important conclusion is that the
inelastic neutron scattering experiments on metallic magnets
permit one to make the process of escaping  the QP very
descriptive.  In this  consideration,  our main emphasis was put
on the aspects important from the point of view of quantum theory
of magnetism, namely, on the dual character of fundamental
"magnetic degrees of freedom". Generally speaking, the fortunate
circumstance in this discussion is the fact that besides the very
general idea of   QP also concrete practical tools are available
in the physics of magnetism, and the combination of these two
approaches is possible in the neutron scattering experiments (
for details see ref.\cite{kuz1}). The approach is very versatile
since it uses the symmetry properties in the most ingenious
fashion. By this consideration an attempt is made to link
phenomenological and quantum theory of magnetism together more
firmly, thus giving a better understanding of the latter.
Finally, to clarify the concept of QP,  we comment on somewhat
resembling mathematical structures which are encountered when one
tries to implement classical dynamic  symmetries in quantum field
theory\cite{marmo}; within these schemes one is trying to fit a
classical description of particles endowed with internal
structures, like spin. However, these analogies, as well as the
elaboration of an adequate mathematical formalism for expression
of the concept of   QP need further studies. Further work is also
necessary for the development of compact criteria appropriate for
the QP occurrence in all applications.


%

\begin{thebibliography}{90}
%
%
\bibitem{kohn}
 W. Kohn,
  {\it Rev. Mod. Phys.} {\bf 71}, S59 (1999).
%
\bibitem{and1}
 P. W. Anderson, {\it Science} {\bf 177},  393 (1972).
%
%
\bibitem{pin1}
 R. B. Laughlin   and D. Pines ,
   {\it Proc. Natl. Acad. Sci. U.S.A. } {\bf 97},  28 (2000).
%
\bibitem{ander}
 P. W. Anderson, {\it Basic Notions of Condensed Matter Physics}
 (Benjamin, N.Y., London, 1984).
%
%
\bibitem{call}
H.B.Callen {\it  Thermodynamics and an Introduction to
Thermostatistics  }   ( J.Wiley and Sons, N.Y., 1985).
%
\bibitem{volov1}
G. E. Volovik, {\it Exotic Properties of Superfluid    $^{3}  He$
}, ( World Scientific, Singapore, 1992).
%
\bibitem{volov2}
 G. E. Volovik,
   {\it Proc. Natl. Acad. Sci. U.S.A. } {\bf 96}, 6042 (1999).
%
%
\bibitem{and2}
 P. W. Anderson,
   {\it Science} {\bf 288}, 480 (2000).
%
%
\bibitem{and3}
  P. W. Anderson,
   {\it Physica C} {\bf 341-348}, 9 (2000).
%
%
\bibitem{pin2}
  D. Pines,
   {\it Physica C} {\bf 341-348}, 59 (2000).
%
%
\bibitem{kuz3}
A. L. Kuzemsky, {\it Communication JINR,} E17-32, Dubna (2000).
%
%
\bibitem{kuz7}
A. L. Kuzemsky, {\em Rivista del Nuovo Cimento}, v.25, n.1, (2002)
1-91.
%
%
\bibitem{pin3}
  Z. Fizk    and D. Pines,
   {\it Nature}  {\bf 394}, 22 (1998).
%
%
\bibitem{mathur}
  N. Mathur,
  {\it Nature} {\bf 400}, 405 (1999).
%
%
\bibitem{kuz1}
A. L. Kuzemsky, in { \it Superconductivity and Strongly
Correlated Electron Systems },
    eds. C. Noce {\em et al.,} ( World Scientific, Singapore, 1994 )  p.346.
%
\bibitem{kuz2}
A. L. Kuzemsky,   {\it Physics of Elementary Particles and Atomic
Nuclei } {\bf 12}, 366 (1981) ;  {\it Sov. J. Part. Nucl.,} {\bf
12}, 146 (1981).
%
%
\bibitem{mathu}
 N. Mathur  {\em et al.,}
   {\it Nature} {\bf 394}, 39 (1998).
%
%
\bibitem{lonz1}
 S. S. Saxena  {\em et al.,}
   {\it Nature} {\bf 406}, 587 ( 2000).
%
%
%
%
\bibitem{allen}
J. W. Allen {\em et al}., {\em Phys.Rev.}  B {\bf 41}, 9013
(1990).
%
\bibitem{steg1}
 F. Steglich {\em et al.,}
   {\it Physica B} {\bf 163}, 44 (1990).
%
%
\bibitem{steg2}
 A. Loidl  {\em et al.,}
   {\it Ann.Physik }  {\bf 1}, 78  (1992).
%
%
\bibitem{tyab}
S. V. Tyablicov, {\em Methods in the Quantum Theory of Magnetism}
(Plenum Press, New York, 1967).
%
\bibitem{her}
C. Herring, {\it Exchange Interactions  among Itinerant Electrons}
(Academic Press, N.Y. 1966).
%
%
\bibitem{mor}
 T. Moriya,
   {\it JMMM }  {\bf 100}, 261  (1991).
%
%
\bibitem{kuze}
 A. L. Kuzemsky,
 {\it Physica  A}  {\bf 267 }, 131 (1999).
%
\bibitem{kuze1}
A. L. Kuzemsky, {\em Nuovo Cimento} {\bf B 109}, 829 (1994).
%
\bibitem{kuze2}
A. L. Kuzemsky, {\em Molecular Phys. Rep.} {\bf 17}, 221 (1997).
%
%
%
\bibitem{smit}
D. A. Smith, {\em J.Physics C:Solid State Phys.}{\bf 1}, 1263
(1968).
%
\bibitem{kiso}
R. Kishore and S. K. Joshi, {\em Phys. Rev.}  B {\bf 2}, 1411
(1970).
%
\bibitem{coq} B. Coqblin
{\em The Electronic Structure of Rare-Earth Metals and Alloys: the
Magnetic Heavy Rare-Earths} (Academic Press, N.Y., London, 1977).
%
\bibitem{kuz4}
 A. L. Kuzemsky,
 {\it Int. J. Mod. Phys.  B} {\bf 13}, 2573 (1999).
%
\bibitem{kuz5}
D. Marvakov, J. Vlahov and A. L. Kuzemsky, {\em J.Physics C:Solid
State Phys.}{\bf 18}, 2871 (1985).
\bibitem{kuz6} D. Marvakov,
A. L. Kuzemsky and J. Vlahov, {\em Physica} {\bf B138}, 129
(1986).
%
\bibitem{kit}
C. Herring and C. Kittel, {\em Phys. Rev.}    {\bf 81}, 869
(1951).
%
\bibitem{low}
S. W. Lowesey {\it  Theory of Neutron Scattering from Condensed
Matter}, vol.1, 2 ( Clarendon Press, Oxford, 1992).
%
\bibitem{marmo}
G. Marmo, G. Morandi  and C. Rubano,  in {\it SYMMETRIES IN
SCIENCE  III},  eds    B. Gruber  and F. Iachello ( Plenum
Publishing,  N. Y.,  1989)  p.243.
%
%
\end{thebibliography}
\end{document}